\documentstyle[aps,multicol,floats,amsfonts,prl]{revtex}
\begin{document}
\draft
\title{Testing Bell's inequality with two-level atoms via 
population spectroscopy}
\author{C. Brif \cite{pa}\ and\ A. Mann \cite{am}}
\address{Department of Physics, Technion---Israel Institute of
Technology, Haifa 32000, Israel}
\maketitle

\begin{abstract}
We propose a feasible experimental scheme, employing methods of 
population spectroscopy with two-level atoms, for a test of Bell's 
inequality for massive particles. The correlation function measured 
in this scheme is the joint atomic $Q$ function. An inequality 
imposed by local realism is violated by any entangled state of a 
pair of atoms.
\end{abstract}


\begin{multicols}{2}

The violation of Bell's inequality \cite{Bell} is a mathematical
expression of the remarkable quantum phenomenon of entanglement.
It may be viewed as a manifestation of the irreconcilability of 
quantum mechanics and ``local realism,'' as was first revealed by 
Einstein, Podolsky, and Rosen (EPR) in their famous gedanken 
experiment \cite{EPR}. Perhaps the most familiar version of the 
EPR experiment, with two spin-half particles in the singlet state, 
is due to Bohm \cite{Bohm}. 

Interest in Bell's inequalities and the EPR ``paradox'' has remained 
amazingly intense during many decades \cite{MR-book,AfSe}.
The violation of Bell's inequalities by entangled quantum states has 
been verified in a number of experiments with photons. Most of them 
used pairs of correlated photons emitted in atomic cascade 
transitions \cite{ACT,Asp82,PDBK85} or produced in parametric 
down-conversion \cite{PDC}. A serious difficulty in 
experiments with photons is the limited detection efficiency. 
Nevertheless, these experiments unequivocally 
confirmed quantum-mechanical predictions in a clear contradiction 
with inequalities imposed by local realism.
A promising experimental scheme was recently proposed \cite{BW99} 
for testing Bell's inequality with entangled two-mode states of the 
quantized light field. In this scheme \cite{BW99}, correlation 
functions are given by the joint two-mode $Q$ function and the 
Wigner function. 
Two decades ago an experiment was made \cite{LaMi76} with pairs of 
spin-correlated protons produced in low-energy proton-proton 
scattering. To the best of our knowledge, this was the only attempt
to demonstrate experimentally the violation of Bell's inequality 
for massive particles.

In the present Letter we propose a feasible experimental scheme for 
testing Bell's inequality with entangled states of a pair of 
two-level atoms, in a situation which is physically very similar and 
mathematically equivalent to the familiar Bohm version of the EPR 
gedanken experiment. We use the fact that the interaction of 
two-level atoms with classical radiation fields is mathematically 
equivalent to the precession of spin-half particles in classical 
magnetic fields. 
These interactions, accompanied by the measurement of the 
atomic-level population, constitute the essence of the Ramsey
population spectroscopy method \cite{Ramsey}. A rearrangement of 
this convenient experimental method was recently proposed for the 
measurement of the spin $Q$ function \cite{BM-unpub}, in line 
with the ideas of Refs.~\cite{Agar98,BM99}. The motivation 
of works \cite{BM-unpub,Agar98,BM99} was to use the measured 
$Q$ function for the reconstruction of the initial quantum state.
Here, we employ the joint $Q$ function of two atoms as the 
correlation function which violates Bell's inequality for 
entangled atomic states. 

If a collection of $N$ two-level atoms (or atomic ions) interacts 
with classical light fields, one can describe this physical 
situation as the interaction of $N$ spin-half particles with 
classical magnetic fields. Denoting by ${\bf S}_r$ the ``spin'' of 
the $r$th particle, one can define the collective spin operators 
\cite{com1}: 
${\bf J} = \sum_{r} {\mathbf{S}}_r$.
The three operators $\{J_{x},J_{y},J_{z}\}$ constitute the su(2)
simple Lie algebra, 
$[J_{k},J_{l}] = i \epsilon_{k l m} J_{m}$.
The Casimir operator ${\bf J}^2$ is a constant times the unit
operator, ${\bf J}^2 = j(j+1) I$, for any unitary irreducible 
representation of the SU(2) group; so the representations are 
labeled by the single index $j$. The representation Hilbert space 
${\cal H}_{j}$ is spanned by the basis 
$\{ |j,m\rangle \}$ ($m=j,j-1,\ldots,-j$):
$J_z |j,m\rangle = m |j,m\rangle$.
The ``cooperative number'' $j$ is a non-negative integer or 
half-integer and takes values 
$\frac{1}{2} N, \frac{1}{2} N - 1,\frac{1}{2} N - 2, \ldots$.
For two atoms, the possible values are $j=1$ (symmetric
triplet states) and $j=0$ (the antisymmetric singlet state). 
We assume that the atoms (ions) are far enough apart so their 
wave functions do not overlap and the direct dipole-dipole coupling 
or other direct interactions between the atoms may be neglected.

In our discussion we will employ concepts of phase space 
and coherent states. For SU(2), the phase space is the unit sphere 
${\Bbb{S}}^2 = {\rm SU}(2) / {\rm U}(1)$, and each coherent state 
\cite{Perel,Gilm} is characterized by a point on the sphere, i.e., 
by a unit vector of the form
${\bf n} = (\sin\theta \cos\phi, \sin\theta \sin\phi, 
\cos\theta)$.
Specifically, the coherent states $|j;{\bf n}\rangle$ are given by
the action of the group elements 
\begin{equation}
\label{eq:oSU2}
g({\bf n}) = e^{- i \phi J_{z} } e^{- i \theta J_{y} }
\end{equation}
on the highest-weight state:
$|j;{\bf n}\rangle = g({\bf n}) |j,j\rangle$.
A useful phase-space distribution function is the $Q$ function:
\begin{equation}
Q({\bf n}) = \langle j;{\bf n} | \rho | j;{\bf n} \rangle ,
\end{equation}
where $\rho$ is the density matrix of the system.

In our scheme, a pair of two-level atoms, prepared initially in 
the state $|\psi\rangle$, is spatially separated \cite{commentN}. 
All subsequent operations (transformations and measurements) are 
made on each atom separately. 
The basic idea of the method can be described as follows. 
Each atom interacts with a classical radiation field of frequency 
close to the atomic transition frequency and/or evolves freely.
This results in a phase-space displacement (rotation) of the
atomic state. Then one measures the population of the upper 
state $|+\rangle$. Of course, such a measurement implies that in 
fact an ensemble of identically prepared systems is used. 
This scheme, which is based on the method routinely employed in 
Ramsey population spectroscopy \cite{Ramsey}, gives values of the
$Q$ function for each atom. Coincidence measurements
for the two atoms give the joint $Q$ function. This function
contains information about quantum correlations between the two 
atoms and can be used in Bell's inequality to distinguish 
between quantum mechanics and local realism.

In the spin description, the magnetic moment 
$\bbox{\mu} = \mu_0 {\mathbf{S}}$ is associated with each particle. 
If a uniform external magnetic field 
${\mathbf{B}}_0 = B_0 \hat{\mathbf{z}}$ 
is applied, the Hamiltonian for each particle is given by
$H_0 = - \bbox{\mu} \cdot {\mathbf{B}}_0 = \hbar \omega_0 S_z$,
where $\hbar \omega_0 = -\mu_0 B_0$ is the separation in energy
between the two levels. (The two eigenstates of $H_0$, denoted by
$|+\rangle$ and $|-\rangle$, are the two atomic levels). 
The corresponding Heisenberg equation is 
$\partial {\bf S}/ \partial t = \bbox{\omega}_0 \times {\bf S}$,
where $\bbox{\omega}_0 = \omega_0 \hat{{\bf z}}$. This spin precession
describes the free evolution of the atom.
In addition, one can apply the so-called clock radiation which is 
a classical field of the form
\begin{equation}
{\mathbf{B}}_{\perp} = B_{\perp} \left( \hat{\mathbf{y}} \cos \omega t
- \hat{\mathbf{x}} \sin \omega t \right) ,
\end{equation}
where $\omega \simeq \omega_0$ and we assume $\omega_0 > 0$.
In the reference frame that rotates at frequency $\omega$, the
spin ${\bf S}$ interacts with the effective field
${\mathbf{B}} = B_{\parallel} \hat{\mathbf{z}} + 
B_{\perp} \hat{\mathbf{y}}$,
where $B_{\parallel} = B_0 (\omega_0 -\omega)/\omega_0$. In the 
rotating frame, the Hamiltonian is 
$H = - \mu_0 {\mathbf{S}} \cdot {\mathbf{B}}$, and the Heisenberg 
equation for $\mathbf{S}$ is 
$\partial {\mathbf{S}}/ \partial t = \bbox{\omega}' \times 
{\mathbf{S}}$,
where $\bbox{\omega}' = (\omega_0 - \omega) \hat{\mathbf{z}}
+ \omega_{\perp} \hat{\mathbf{y}}$ and 
$\omega_{\perp} = -\mu_0 B_{\perp}/\hbar$. We do not use any 
special notation for the rotating frame because the population of 
the upper state $|+\rangle$, which will be measured, is the same 
in both frames.

In the Ramsey method there are two kinds of interactions. In the
first kind, $B_{\perp}$ is nonzero and constant during the time 
interval $T_{\theta}$. During this period
${\mathbf{B}} \simeq B_{\perp} \hat{\mathbf{y}}$, where it is
assumed that $|B_{\perp}| \gg |B_{\parallel}|$, i.e., 
$|\omega_{\perp}| \gg |\omega_0 - \omega|$. (This assumption means 
that the free evolution can be neglected because it is very slow 
compared to the precession induced by the clock radiation). 
Therefore, in the rotating frame, $\mathbf{S}$ rotates around the 
$\hat{\mathbf{y}}$ axis by the angle 
$\theta = \omega_{\perp} T_{\theta}$.
The second kind is just the free evolution. During the time 
interval $T_{\phi}$ the clock field $B_{\perp}$ is zero, so 
${\mathbf{B}} = B_{\parallel} \hat{\mathbf{z}}$, and 
$\mathbf{S}$ rotates around the $\hat{\mathbf{z}}$ axis by the 
angle $\phi = (\omega_0 - \omega) T_{\phi}$. 

Now, assume that the $r$th atom ($r=1,2$) first evolves freely, 
resulting in a rotation around the $\hat{\mathbf{z}}$ axis by 
$-\phi_r$, and then interacts with the clock radiation,
resulting in a rotation around the $\hat{\mathbf{y}}$ axis by
$-\theta_r$. The overall transformation performed on the whole
system of two atoms is the phase-space displacement (rotation):
\[
|\psi( {\bf n}_1 , {\bf n}_2 )\rangle = \prod_{r=1,2} 
e^{ i \theta_r S_{r y}} e^{ i \phi_r S_{r z}} |\psi\rangle
= g_{1}^{\dagger}( {\bf n}_1 ) g_{2}^{\dagger}( {\bf n}_2 ) 
|\psi\rangle . \]

The last step of the experimental procedure is measurement of the 
upper-level populations of the atoms. This measurement can be made 
with almost unit efficiency using the technique of ``quantum jumps'' 
\cite{qjumps}. The whole procedure should be repeated with many 
identically prepared systems.
Define $q_r^{(l)}$ ($r=1,2$) which is $1$ if the $r$th atom is found 
in the upper state $|+\rangle$ in the $l$th experiment, and $0$ 
otherwise. Repeating the described procedure with $L$ identical 
systems ($L \gg 1$), one determines the probability to find the 
$r$th atom in the upper state,
$Q_r = L^{-1} \sum_{l = 1}^{L} q_r^{(l)}$. 
For the displaced state $|\psi( {\bf n}_1 , {\bf n}_2 )\rangle$, 
the probabilities $Q_r$ are
\begin{eqnarray}
Q_r ( {\bf n}_r ) & = & {}_r \langle +| {\rm Tr}_{p \neq r}
\left\{ |\psi( {\bf n}_1 , {\bf n}_2 ) \rangle \langle 
\psi( {\bf n}_1 , {\bf n}_2 ) | \right\} |+\rangle_r \nonumber \\
& = & {}_r \langle {\bf n}_r | \rho_r | {\bf n}_r \rangle_r ,
\hspace{6mm} r = 1,2 .
\end{eqnarray}
Here, $\rho_r = {\rm Tr}_{p \neq r} \left\{ |\psi \rangle 
\langle \psi | \right\}$ is the reduced density matrix of the $r$th 
atom, and $| {\bf n}_r \rangle_r = g_r ( {\bf n}_r ) |+\rangle_r$
is the atom's coherent state, 
\begin{equation}
| {\bf n} \rangle = \cos(\theta/2) e^{- i \phi/2} |+\rangle
+ \sin(\theta/2) e^{ i \phi/2} |-\rangle .
\end{equation} 
Hence, $Q_r ( {\bf n}_r )$ is just the $Q$ function for the 
$r$th atom.

In a similar way, the probability to find both atoms in the upper 
state, determined in experiments with $L$ identical systems, is 
$Q_{12} = L^{-1} \sum_{l = 1}^{L} q_{1}^{(l)} q_{2}^{(l)}$. 
For the displaced state $|\psi( {\bf n}_1 , {\bf n}_2 )\rangle$, 
this probability is 
\begin{eqnarray}
Q_{12} ( {\bf n}_1 , {\bf n}_2 ) & = & \left| {}_1 \! \langle +|\,
{}_2 \! \langle + | \psi( {\bf n}_1 , {\bf n}_2 ) \rangle \right|^2
\nonumber \\
& = & \left| {}_1 \! \langle {\bf n}_1 |\,
{}_2 \! \langle {\bf n}_2 | \psi \rangle \right|^2 .
\end{eqnarray}
This is just the joint $Q$ function for the system of two atoms.
Therefore, ``rotations'' performed on the atoms followed by the 
measurement of the upper-state populations (procedures routinely
employed in population spectroscopy) allow one to measure values
of the $Q$ functions at any phase-space point.
Note that the free evolution (represented by a rotation around the
$\hat{\mathbf{z}}$ axis) does not affect the level populations, if 
it is not followed by a rotation around an axis different from 
$\hat{\mathbf{z}}$.

The dichotomic observables $q_1$ and $q_2$ and the corresponding
joint probability $Q_{12}$ (the correlation function) are just what 
one needs to formulate a Bell inequality. Define the combination of 
the Clauser-Horne type \cite{ClHo74}
\begin{eqnarray}
\Gamma & = & Q_{12} (0,0) + Q_{12} ({\bf n},0) + 
Q_{12} (0,{\bf n}') \nonumber \\
& & - Q_{12}({\bf n},{\bf n}') - Q_1 (0) - Q_2 (0) .
  \label{eq:gamma}
\end{eqnarray}
Local realism requires that $\Gamma$ must satisfy the inequality 
$-1 \leq \Gamma \leq 0$. If two atoms are prepared in an entangled 
state, quantum mechanics predicts that a measurement performed on 
one of them ``affects'' the other. This contradicts local realism 
and is expressed mathematically by the violation of Bell's 
inequality. Consider two atoms prepared in the entangled state 
\begin{equation}
\label{eq:ustate}
| u(\varphi) \rangle = 2^{-1/2} \left( |+\rangle_1 |-\rangle_2 
+ e^{ i \varphi} |-\rangle_1 |+\rangle_2 \right) .
\end{equation}
This is the singlet state $|j=0,m=0\rangle$ for $\varphi = \pi$ 
and the triplet state $|j=1,m=0\rangle$ for $\varphi = 0$. 
The calculation of the individual and joint atomic $Q$ functions 
is rather straightforward. For $\theta = \theta'$, we find 
\begin{equation}
\label{eq:gamma-s1}
\Gamma = \sin^2 \frac{\theta}{2} - \frac{1}{2} \sin^2 \theta 
\cos^2 \frac{\phi-\phi'-\varphi}{2} - 1 .
\end{equation}
The maximum value of $\Gamma$ is $0$, so we look for a minimum.
For $\phi - \phi' = \varphi$, we get
$\Gamma = - 1 - \frac{1}{2} (\cos \theta - \cos^2 \theta)$.
This function reaches its minimum $\Gamma = -9/8 = -1.125$ for 
$\theta = \pi/3$. This clearly violates the limit imposed by 
local realism. 

If one considers two light modes $a$ and $b$ which share a single
photon, the description is very similar to the case of two 
spin-half particles. The singlet state then takes the form
$| u(\pi) \rangle = 2^{-1/2} \left( |1\rangle_a |0\rangle_b 
- |0\rangle_a |1\rangle_b \right)$, where the one-photon state 
$|1\rangle$ is identified with the upper spin state $|+\rangle$ 
and the vacuum state $|0\rangle$ with the lower spin state
$|-\rangle$. Peres \cite{Per95} proposed the following 
experiment: one observer measures the projections of 
$| u(\pi) \rangle$ on the states $|1\rangle_a$ and
$|\omega_+ \rangle_a = \frac{1}{2} (\sqrt{3} |1\rangle_a +
|0\rangle_a )$ of the mode $a$ and the other observer
measures the projections of $| u(\pi) \rangle$ on the states 
$|1\rangle_b$ and $|\omega_- \rangle_b = \frac{1}{2} 
(\sqrt{3} |1\rangle_b - |0\rangle_b )$ of the mode $b$.
Using these measurements, one can construct the Clauser-Horne
combination $\Gamma$ which violates Bell's inequality,
$\Gamma = -9/8$. While in the context of the Fock spaces of
light modes the states $|\omega_+ \rangle_a$ and 
$|\omega_- \rangle_b$ look somewhat contrived, in the context of 
two-level atoms they are easily recognized as the atomic coherent 
states $|{\bf n}\rangle_1$ and $|{\bf n}'\rangle_2$ with 
$\theta = \theta' = \pi/3$ and $\phi - \phi' = \pi$.
As we just found above, projections on these coherent states
can be used to obtain the maximum violation of local realism,
$\Gamma = -9/8$. It is important to emphasize that in the context
of two-level atoms these projections on coherent states are
physically meaningful (and experimentally feasible), as they
can be measured by means of appropriate operations of population
spectroscopy.

Another possible entangled state of two atoms is
\begin{equation}
\label{eq:vstate}
|v(\varphi)\rangle = 2^{-1/2} \left( |+\rangle_1 |+\rangle_2 + 
e^{ i \varphi} |-\rangle_1 |-\rangle_2 \right) .
\end{equation}
This state belongs to the triplet subspace ($j = 1$).
For $\theta = \theta'$, we find
\begin{equation}
\label{eq:gamma-xi1}
\Gamma = \frac{1}{2} \left( \cos\theta - \cos^2 \theta - 
\sin^2\theta \cos^2 \frac{\phi+\phi'-\varphi}{2} \right) .
\end{equation}
The minimum value of $\Gamma$ is $-1$, so we look for a maximum.
For $\phi + \phi' - \varphi = \pi$, we get
$\Gamma = \frac{1}{2} (\cos \theta - \cos^2 \theta)$.
This function reaches its maximum $\Gamma = 1/8  = 0.125$ for 
$\theta = \pi/3$. Once again, an entangled state violates the 
limit imposed by local realism.

It should be noted that enormous theoretical and experimental 
progress has been made during the last few years in the generation 
of entangled atomic states. 
The use of trapped atomic ions was proposed in a number of recent 
works \cite{ME} to generate the state (\ref{eq:vstate}) 
and its multiparticle generalizations,
$2^{-1/2} ( |j,j\rangle + e^{ i \varphi} |j,-j\rangle )$.
Deterministic entanglement of two trapped ions was reported in
\cite{Tur98}. Entangled states created in this experiment are 
similar to those of Eq.~(\ref{eq:ustate}). 
Another method for the deterministic generation of entangled
states of two trapped ions was recently proposed in \cite{Sol99}.
Experimental preparation of entangled states of pairs of Rydberg 
atoms in a cavity was reported in \cite{Hag97}. A method was also 
proposed \cite{Bri99} to entangle neutral atoms via cold 
controlled collisions. Maximally entangled states of three spins 
were created in a recent experiment \cite{Laf98} using nuclear 
magnetic resonance.

It is known that factorizable quantum states (i.e., states with
no entanglement between the two subsystems) satisfy Bell's 
inequality. Consider a system of $N$ two-level atoms with 
``cooperative number'' $j$ which is split into two subsystems
with cooperative numbers $j_1$ and $j_2$ such that 
$j = j_1 + j_2$ (e.g., separation of two atoms in a triplet-basis
state with $j=1$). 
Then the highest state and the lowest state factorize:
$|j,\pm j\rangle = |j_1 , \pm j_1 \rangle_1 
|j_2 , \pm j_2 \rangle_2$.
More generally, if the whole system is in the
coherent state $|j,{\bf n} \rangle$, then it also factorizes
\cite{BMR98}:
\begin{eqnarray}
|j,{\bf n}\rangle = g({\bf n}) |j,j\rangle & = &
g_1 ({\bf n}) g_2 ({\bf n}) |j_1 , j_1 \rangle_1 
|j_2 , j_2 \rangle_2 \nonumber \\
& = & |j_1 ,{\bf n}\rangle_1 |j_2 ,{\bf n}\rangle_2 .
\end{eqnarray}
Here, we used the rule for the addition of angular momenta,
${\bf J} = {\bf J}_1 + {\bf J}_2$, to obtain
$g({\bf n}) = g_1 ({\bf n}) g_2 ({\bf n})$. 
As a result, the correlation function factorizes as well:
\begin{equation}
Q_{12} ( {\bf n}_1 , {\bf n}_2 ) = 
\left( \frac{ 1 + {\bf n} \cdot {\bf n}_1 }{2} \right)^{ 2 j_1 }
\left( \frac{ 1 + {\bf n} \cdot {\bf n}_2 }{2} \right)^{ 2 j_2 } .
\end{equation}
Therefore, a system of two-level atoms prepared in the coherent state
never violates Bell's inequality (provided that $j = j_1 + j_2$).
Moreover, it can be proved \cite{BMR98} that atomic coherent states 
are the only pure states which 
factorize upon splitting and thereby never violate Bell's inequality 
\cite{AFLP-MRS}. (Note that the highest state and the lowest state 
are particular cases of coherent states.)

On the other hand, any pure entangled state (i.e., any pure state 
which does not factorize) can always, by a suitable
choice of local measurement apparatus, be made to violate Bell's
inequality \cite{Per78,Gis91}. In definition (\ref{eq:gamma}) 
the measurement apparatus is partially fixed, because the measured
observables were chosen to be the upper-level populations of the
atoms. However, we still have to choose displacement parameters 
(rotation angles), and this freedom is very essential. 
In fact, \emph{any pure entangled state of two atoms violates the 
inequality} $-1 \leq \Gamma \leq 0$. 
This is intuitively clear because of the following argument. 
The measurement of the upper-level population of an atom gives the 
projection of the associated spin ${\bf S}$ on the $\hat{{\bf z}}$ 
axis. However, the projection on any axis can be measured in this 
way if one forestalls the population detection with an appropriate 
rotation. This means that in fact we are free in the 
choice of the dichotomic variables measured by local apparatus. 

For a more rigorous proof, let us consider the state
$|\eta\rangle = \cos \vartheta |+\rangle_1 |+\rangle_2 + 
\sin \vartheta\, e^{ i \varphi} |-\rangle_1 |-\rangle_2$.
An arbitrary pure state $|\psi\rangle$ can be written in the 
Schmidt decomposition as 
$|\psi\rangle = \cos \vartheta |\xi_+ \rangle_1 
|\zeta_+ \rangle_2 + \sin \vartheta\, e^{ i \varphi} 
|\xi_- \rangle_1 |\zeta_- \rangle_2$,
where $|\xi_{\pm}\rangle_1$ and $|\zeta_{\pm}\rangle_2$
are some orthonormal bases for the first and the second 
atom, respectively. The state is entangled if 
$\vartheta \neq \frac{1}{2} n \pi$ $(n=0,\pm 1, \ldots)$. 
Clearly, $|\psi\rangle$ can be obtained
from $|\eta\rangle$ (with the same $\vartheta$ and $\varphi$) by 
rotations of the single-atom bases: 
$|\psi\rangle = g_1 ({\bf n}) g_2 ({\bf m}) |\eta\rangle$
(the degree of entanglement, dependent on $\vartheta$, is conserved 
under these local transformations). 
Therefore, the joint $Q$ function of an arbitrary state 
$|\psi\rangle$ at a point $( {\bf n}_1 , {\bf n}_2 )$ is equal to 
the joint $Q$ function of $|\eta\rangle$ at some other point. 
We are able to show that the state $|\eta\rangle$ 
violates the inequality $-1 \leq \Gamma \leq 0$ as long as it is
entangled (the amount of violation depends on the degree of 
entanglement and has a maximum $\Gamma = 0.125$ for 
$\vartheta = \pi/4$). 
Consequently, any pure entangled state violates this 
inequality.

In conclusion, we presented an efficient experimental method to
test the violation of Bell's inequality by entangled states of
two-level atoms. Our method employs the convenient technique of
population spectroscopy which is characterized by very high
measurement accuracy (the detection efficiency is almost unity). 
The correlation function measured by this method is just the joint 
atomic $Q$ function. The corresponding Clauser-Horne combination 
$\Gamma$ violates the inequality $-1 \leq \Gamma \leq 0$, imposed by 
local realism, for any entangled state of a pair of atoms. Maximal 
violations are found to be $\Gamma = -1.125$ and $\Gamma = 0.125$.

C.B. was supported by the Lester Deutsch Fund and by the Institute 
of Theoretical Physics at the Department of Physics at the Technion.
A.M. was supported by the Fund for Promotion of Research 
at the Technion and by the Technion VPR Fund.

\end{multicols}


\begin{references}
\bibitem{commentN} We focus on the fundamental case of a pair of 
        atoms, but in principle the method may be generalized for 
        systems consisting of more than two atoms.
\end{references}
\end{document}